\documentclass[11pt]{article}
\usepackage{graphicx}

\usepackage{picture}
\usepackage{hyperref}      
\usepackage{picture}

\newcommand\pubnumber{SLAC--PUB--15859}
\newcommand\pubdate{December, 2013}

\def\Title#1{\begin{center} {\Large #1 } \end{center}}
\def\Author#1{\begin{center}{ \sc #1} \end{center}}

\newcommand\pubblock{\rightline{\begin{tabular}{l} \pubnumber\\
         \pubdate \end{tabular}}}
\newenvironment{Abstract}{\begin{quotation} \begin{center}
                       ABSTRACT
     \end{center}\bigskip  }{\end{quotation}}
\newenvironment{Presented}{\begin{quotation} \begin{center} 
            CONTRIBUTED TO\end{center}\bigskip 
      \begin{center}\begin{large}}{\end{large}\end{center} \end{quotation}}

\def\Acknowledgements{\bigskip  \bigskip \begin{center} \begin{large}
             \bf ACKNOWLEDGEMENTS \end{large}\end{center}}

\begin{document}
\begin{titlepage}
\pubblock

\vfill
\Title{Benefits to the U.S.  from  Physicists Working at Accelerators Overseas}
\vfill
\Author{Jacob Anderson, Raymond Brock, Yuri Gershtein, Nicholas Hadley, Michael
  Harrison, Meenakshi Narain, 
Jason Nielsen, Fred Olness, Bjoern Penning,  Michael Peskin, Eric
Prebys, Marc Ross,  
Salvatore Rappoccio, Abraham Seiden, Ryszard Stroynowski}
\vfill
\begin{Abstract}
We illustrate benefits to the U.S.  economy and technological
infrastructure of U.S.  participation in accelerators overseas.  We
discuss
contributions to experimental hardware and analysis and to 
accelerator technology and components, and benefits stemming 
from the involvement of U.S.  students and postdoctoral fellows in 
global scientific collaborations.

\end{Abstract}
\vfill
\begin{Presented}
Snowmass 2013 Electronic Proceedings \\
Community Summer Study, Minneapolis, MN \\
July 29 -- August 6, 20123\end{Presented}
\vfill
\end{titlepage}
\def\thefootnote{\fnsymbol{footnote}}
\setcounter{footnote}{0}

\tableofcontents

\newpage

\section{Introduction}

High Energy Physics is increasingly done in large collaborations at particular sites around the world --- either because the location is itself special (such as in Antarctica or at high altitude or underground locations) or because the science specifies a facility requiring an underwriting that cannot be borne solely by any single nation. The nature of these global collaborations is often misunderstood.   The discovery of the Higgs boson at the Large Hadron Collider at CERN gives 
a recent example.  On  July 4, 2012, this discovery was front-page news around the world.  U.S.  physicists' contributions were central to this breakthrough.  Yet news reports on the Higgs boson often included the sentiment that this was a lost opportunity for the United States.  For example, a March 4 headline in the New York Times read: ``Particle Physicists in the U.S. Worry About Being Left Behind''~\cite{PPLB}.

This sentiment came as a surprise to those of us actively working in particle physics. What we see in our universities and national labs is vibrant real-time participation in the experiments at the LHC  and aggressive planning for the next twenty years of research there.  We see large numbers of young scientists who are caught up in the excitement of this program.  It is true that the particle detectors are located in Geneva, Switzerland, but the LHC experiments and many aspects of the accelerator construction are globalized endeavors including critical U.S.  involvement.

The purpose of this report is to clarify the concepts involved in discussing U.S. involvement 
in global scientific projects. Our examples will be drawn from U.S. collaboration in the 
LHC and related global collaborations involving particle accelerators. 
 We hope that this account  will be helpful to our particle 
physics colleagues in framing their ideas on this subject.  We hope that it will also be 
useful to our colleagues in other fields  who 
wonder about just what goes on in  the large particle physics 
collaborations and how that benefits U.S.  science. 

Discussions today about the future of the U.S. high energy physics program often pose the 
question:  What are the advantages to the U.S. economy of supporting research 
carried out at accelerators overseas?  It is not possible to address  this question without understanding 
how global  scientific collaborations actually operate. 
As a framework for analyzing this question,  we 
characterize the support of basic science by U.S. taxpayers as an investment returning 
rewards of three types:
\begin{enumerate} 
\item It insures continuing U.S.  leadership in the intellectual challenge of making fundamental
discoveries about nature.
\item It develops equipment of high technological capability, benefitting the U.S.  economy both in its construction and in the development of technical tools needed in that construction. We call this
``Innovation-Transfer.''
\item It trains young scientists to develop methods that are beyond the current state of the art, advancing science and industry across a broad front. HEP students learn to solve problems by inventing the tools. We call this ``Imagination-Transfer.''
 \end{enumerate}
In this report, we will demonstrate with examples how the  U.S. participation in the LHC and related 
global accelerator projects fulfills all three of these goals.   The logic of our argument extends to any 
other conventionally structured  international venture in particle physics.

Make no mistake, working overseas is indeed different. We travel often and over longer distances, have virtual meetings at odd hours over eight or more time zones, and write and sign papers together with vast numbers of collaborators.  However, it is also important to emphasize the ways in which  physics is done in the same way now as before. The nature of human interactions means that most physics analysis or detector performance groups of LHC physicists are not larger than a few dozen people.   It is still true that the the basic unit of discovery is ultimately a professor, a student, and a lab bench, blackboard, or computer monitor.

The report is organized as follows:  In Section 2, we discuss the structure of global science
collaborations.  Using the LHC collaborations as an example, we discuss how U.S. physicists
play leadership roles. In Section 3, we provide examples of innovation-transfer stemming from 
U.S. participation in LHC, both in the detectors and in the accelerator.  In Section 4, we 
present some example of imagination-transfer from the LHC experience.
In Section 5, we discuss the role that Fermilab can play as a facilitator of U.S.  involvement in high-energy physics projects abroad.  Section 6 gives our conclusions.

\section{U.S. Leadership in Fundamental Science}

``Big Science'' is a modern-day phrase to describe scientific projects that are so expensive that they require centralized national support and oversight.  There is nothing
new about Big Science.  The Babylonian and Mayan observatories and Tycho Brahe's 
Uraniborg provide examples. However, in the 21st century, the ease of global communication
and travel and scale of our scientific ambitions have brought Big Science to a new level.
In many areas of science today, the largest projects are such that active investment and committed scientific collaboration  from multiple nations is required in order to accomplish the project's scientific goals.

\subsection{Today's Big Science Scale}

Particle physics has been a leader in the drive toward Big Science projects of very large scale.
Our machines have evolved from cyclotrons that could be held in one's hand, to those housed in the basements of many university laboratories, to particle accelerators at the scale of laboratory and university campuses, to, now, those of the size of whole counties.  The growth of these facilities has been accompanied by a consolidation of resources, with the result that leading accelerators are 
 hosted by only a few international laboratories.  This path has been the natural result of following the physics of the smallest constituents of matter 
and the most intricate fundamental forces among them.
This has led us to develop probes at increasingly short distances, requiring increasingly higher 
energies.

The largest accelerator in use today is the Large Hadron Collider at CERN.   The cost of constructing the LHC, including personnel and materials is estimated by CERN as CHF 5B\footnote{In recent years the relative value of  CHF and \$ has been about 1:1.}~\cite{LHCcost}. The cost of the accelerator was largely borne by CERN
itself, from the subscriptions of the 20 CERN member states, with the U.S., Japan, and other 
non-member nations contributing specialized in-kind contributions.   The four large 
detectors ALICE, LHCb, ATLAS, and CMS were constructed by world-wide consortia.  The CERN contribution to their cost was about 
CHF 1.362B including salaries~\cite{LHCcost}.  The CERN contribution ranged from 14\% to 20\% of the total, with the rest coming from the nations whose physicists participate
in the experiments.   The annual expense of running the LHC is more than CHF 300M \cite{cernbudget}.   The full cost of the LHC project then runs above \$10B: this is Big Science by any measure.

But today examples of projects with a similar scale of cost can be found in many disciplines.
 In  materials physics, the Basic Energy Sciences division of the Department of Energy operates a nationally-distributed complex of four major synchrotron light sources plus the Spallation Neutron Source and the LCLS X-ray laser, for a total investment of about \$4B.  European labs for neutron and laser beams have 
been constructed at similar cost. These began as individual national ventures but have evolved 
into European collaborative facilities such as the European Spallation Source ESS and the European XFEL.
In astronomy, much important data now come from satellites with unique capabilities designed to observe in  particular wavebands or from ground-based interferometric optical and radio observatories, often at unusual geographic locations.  Even in the visible region, the unparalleled resolution of the Hubble Space Telescope has opened new observational opportunities.
In 2011 dollars, the Hubble Program, including upgrades, has cost about \$8B.  Our nation has now promised the next-generation optical facility,  the James Webb Space Telescope, for which the projected cost for the initial satellite is  \$8B.  In biology, the Human Genome Project, an endeavor mainly supported by the U.S.  through a partnership of the  Department of Energy and the National Institute of Health,  had a cost at this scale, roughly \$4.6B in current-year dollars. The list goes on: ITER is of a scale of \$18B.   The Apollo program and the International Space Station---arguably not basic science projects---each cost more than \$100B
in current dollars. Big Data projects from climate and ecosystem modeling to biological sequencing now approach the \$B threshold.

There is no turning back from Big Science.  These large facilities have enabled scientific discoveries that would not have been possible in projects of smaller scale.  In particle
physics, the LHC, with its associated large physics organizations and global participation, was an absolute requirement for the discovery of  the Higgs boson.   Looking forward, future facilities of the same scale will be needed to produce other new particles at the TeV energy scale for detailed study in the laboratory.    It is thus increasingly important to understand how large scientific projects are managed, appreciate how their benefits should be understood, and to recognize the forms of intellectual leadership in projects of this scale.

\subsection{Sharing of effort in large science projects}

One might imagine that a multi-\$B project is organized as  a tightly structured
hierarchical tree of strictly accountable managers with corporate-like focus on responsibility and productivity.  However, large particle physics projects have never operated in this way.
Our traditions grew out of a culture of small experimental groups in which all members 
exercised their creativity and decisions were taken by consensus.   To as great an extent as
possible, the current structure of large particle physics collaborations preserves these 
ideals within a democratic management structure.

 There is an overall scientific leadership, with  a single Spokesperson or, in some cases,  two co-Spokespersons,  who have the final authority on major decisions. These officials
 are elected by the membership of the collaboration and then serve for fixed terms. They must have the trust of their collaborators and also, importantly, that of the host lab and other major laboratory partners.    Other elected executives
provide the overall coordination of the physics analyses and detector maintenance and improvement.

Below these executives are working groups organized under  ``conveners.''    There are dozens of
such groups and thus dozens of conveners who guide either physics analysis or performance evaluation of the detector components. There are also committees which regulate publication, construction, and operations strategies. These positions are usually also elected, although some are appointed. 

There are also many groups charged with constructing and/or upgrading the apparatus.  In these
groups, the  leadership is chosen through competitive proposals  to produce a proposed piece of apparatus. Leaders here are most visible as the PIs of equipment and construction grants.  Computing infrastructure, which is essential to the operation of these experiments, is treated as a part of the technical capability.

It is noteworthy that the host country or host laboratory has no special privilege to appoint to 
any of these positions.

In both of the major LHC experiments ATLAS and CMS,  U.S. physicists
 play a substantial role in this 
administrative structure.    Table~\ref{tab:phys}~\cite{LHCauthors} shows the 
census of the two experiments. In both collaborations, the U.S. is the largest single national partner.  U.S. physicists make up roughly a quarter of the total number of collaborators.
Table~\ref{tab:lead} shows the executive  positions held by U.S. physicists. In addition to the spokesperson-level management, U.S. physicists in both collaborations are or have been leaders of Physics, Upgrade, Computing, and collaboration-wide advisory bodies..
It is important to note that U.S. physicists play a role in the convenerships and 
group leader positions quite commensurate with our substantial participation
overall  in the collaborations.

\begin{table}[htp]
\centering
\begin{tabular}{lrrr}
\hline\hline
Category			& Authors & U.S. & Percentage	\\ \hline
ATLAS authors		& 2900 		& 583 	& 20\%  \\
ATLAS graduate students & 1000		&214	&21\% \\
ATLAS nations 		& 37 		& 1 		& 3\% \\
ATLAS institutions 	& 194 		& 44		& 23\% \\
\hline \hline
Total CMS authors	& 2500 		& 678	&  27\%  \\
CMS graduate students & 600	& 247	& 41\% \\
CMS nations 		& 40			& 1		&  3\% \\
CMS institutions 	& 183 		& 49		& 27\%  \\
\hline\hline
\end{tabular}
\caption{U.S. physicists in the ATLAS and CMS collaborations.}
\label{tab:phys} 
\end{table}

\begin{table}[htp]
\centering
\begin{tabular}{l|cc}
\hline\hline
ATLAS executive positions  \\ \hline
 \qquad Deputy Spokesperson &    Beate Heinemann (UC Berkeley)  \\  \hline
       &  U.S. fraction    &  percentage  \\ \hline 
 ATLAS physics group conveners   &   3/16	& 19 \% \\
ATLAS detector performance conveners 	& 8/20 	& 40\% \\ \hline \hline
CMS  executive positions  \\ \hline
 \qquad Spokesperson &   Joseph Incandela (UCSB)  \\  \hline
       &  U.S. fraction    &  percentage  \\ \hline 
CMS physics group conveners   &  8/18	& 44\% \\
CMS detector performance conveners 	& 3/14 	& 21\% \\ \hline
\end{tabular}
\caption{U.S. physicists in leadership positions  in the ATLAS and CMS experiments. }
\label{tab:lead} 
\end{table}

\subsection{Sharing international responsibility for large projects}

Large science projects bring both costs and benefits.   Nations that bear a major fraction
of the costs are also in the position to purchase more from domestic industry and to 
be more closely involved with industry in technology development.   There are also special
benefits that come if a nation or region's lab is the host of a global project.
  One that is obvious for the LHC experiments is that CERN manages the relations of its experiments with the press and expects well-deserved credit for the physics results from the LHC.

Each nation that participates in Big Science has the opportunity to choose the projects that it will
host and the projects for which it will collaborate in other regions.  These are political decisions.  They are also shaped by regional
culture.  The U.S. has been and remains the host of the largest-scale space missions.
 The special historic status of CERN and its importance as a symbol of European cooperation
has driven European science investment toward high-energy physics.  In Asia, the stronger cultural importance of physical science and the demands of national industrial policy have also favored high-energy physics investments. 

For us as working physicists, the most important consideration is that we should be able to pursue what we, in our professional evaluation, consider the most important problems in fundamental science.  Those of us who are university professors also see an importance in preserving this opportunity for our students.  

The United States research university system has no equal in the world.   Its success comes from the unique way in which our system nurtures creativity and allows new ideas from ambitious young people to drive scientific projects forward.   In an era of global collaboration, one should ask what advantage U.S.  universities have over others in the world.  The answer can only be that our universities are engaged in the most important problems of science and that they give  our students the best opportunity to make original contributions to these areas of research.

It is therefore crucial to understand that U.S.  participation in major physics projects hosted abroad preserves this opportunity.  It is equally important to understand that this participation preserves the other benefits that society asks from its investment in fundamental science.  Only on this basis can we correctly compare the merits of  projects based in the U.S.  and projects with investments of similar size at accelerators  hosted abroad.

\section{Particle Physics Innovation-Transfer}

In many sciences, the technology is commercially developed and purchased by scientists for use in their labs.  Chemists and biologists purchase NMR, optical, and infrared spectrometers along with their laboratory flasks and desktop computers.  Most fields of physics, however, require unique
experimental equipment that must be designed and built by the scientists themselves.  Particle
physics provides the most extreme examples of this, requiring both accelerators and massive
detectors, each component of which must exceed the state of the art in some way.  The design,
construction, and maintenance of these devices is an essential part of the art of 
doing science in our 
field.

This is both a burden and a benefit. The burden is that our scientific teams must continually create new technological solutions to address their scientific goals.  The benefit is that our system of work and our system of education fosters the skills needed to create new technologies. Inevitably, these solutions spin off into the marketplace, often  in ways which were largely unpredictable during their germination. This is true both for accelerator and detector components and for the
electronics and computing required in order develop and use them.

Accelerators designed for scientific applications are a striking
example of this evolution.  The world has needed only a small number
of multi-GeV proton and electron accelerators and indeed, our global
community works well today with only one LHC.  Yet there are more than
30,000 accelerators in use in the world, more than 100 for every one
used in basic research~\cite{countacc}. Almost 70 companies produce a
thousand accelerators a year which provide essential manufacturing
stages in an astonishingly wide range of industries: medicine,  chip
manufacture,  food sterilization, and  defense and security
applications~\cite{man70}. The technical innovations that make their
way to industrial applications and, importantly, the accelerator
scientists who migrate to this growing particle beam industry, both
have their roots in the original need to solve scientific problems.
It's not an exaggeration to conclude that Big Science problems and the
international will to solve them with individually designed
accelerators enable and even create hundred-billion-dollar industries.

It is thus directly relevant to U.S. technological development that U.S. scientists collaborate 
in frontier accelerators and in the detector systems that use them, even if these accelerators 
are located abroad.   In this section, we clarify how the work of U.S. scientists in these
collaboration  proceeds.  We will explain the mode of these collaborations, where the work is done, 
and where the money for these developments is spent.

\subsection{Contributions to Detectors}

We first discuss the U.S.  contributions to the LHC detectors.  This contribution is substantial, approximately \$200M for materials and a larger amount for the salaries of U.S.  physicists involved in the construction.  Almost all of this money has been spent in the U.S.,  with benefits equivalent to those of any U.S.-hosted physics project.  

Table~\ref{tab:detect} lists the resources that were expended \textit{here in the U.S.} in order to build the ATLAS and CMS detectors. In all cases the equipment that was fabricated was done with U.S. manufacturers or at U.S. laboratories with staff and student labor. In all cases, the salaries that are listed were paid to U.S. scientists, engineers, and students. This money stayed in the U.S. and educated students and industry in cutting-edge manufacturing.   The spending is distributed across the U.S., supporting projects based at 
all of the 93 universities and laboratories indicated in Table~\ref{tab:phys}. 

\begin{table}[tbp]
\centering
\begin{tabular}{lr}
\hline\hline
Source			& Equipment \& Personnel				\\ 
\hline
U.S. Department of Energy, Office of Science	& \$250M  	\\
National Science Foundation, Division of Physics  & \$81M		\\
\hline
Total									& \$331M	\\
\hline
Total to U.S. ATLAS		& \$164M    \\
Total to U.S. CMS  		& \$167M	\\
\hline\hline
\end{tabular}
\caption{U.S. resources spent for ATLAS and CMS detector construction~\cite{resources}}
\label{tab:detect} 
\end{table}

\subsubsection{CMS endcap}

\begin{figure}[p] 
\begin{center} 
\includegraphics[height=6.0in]{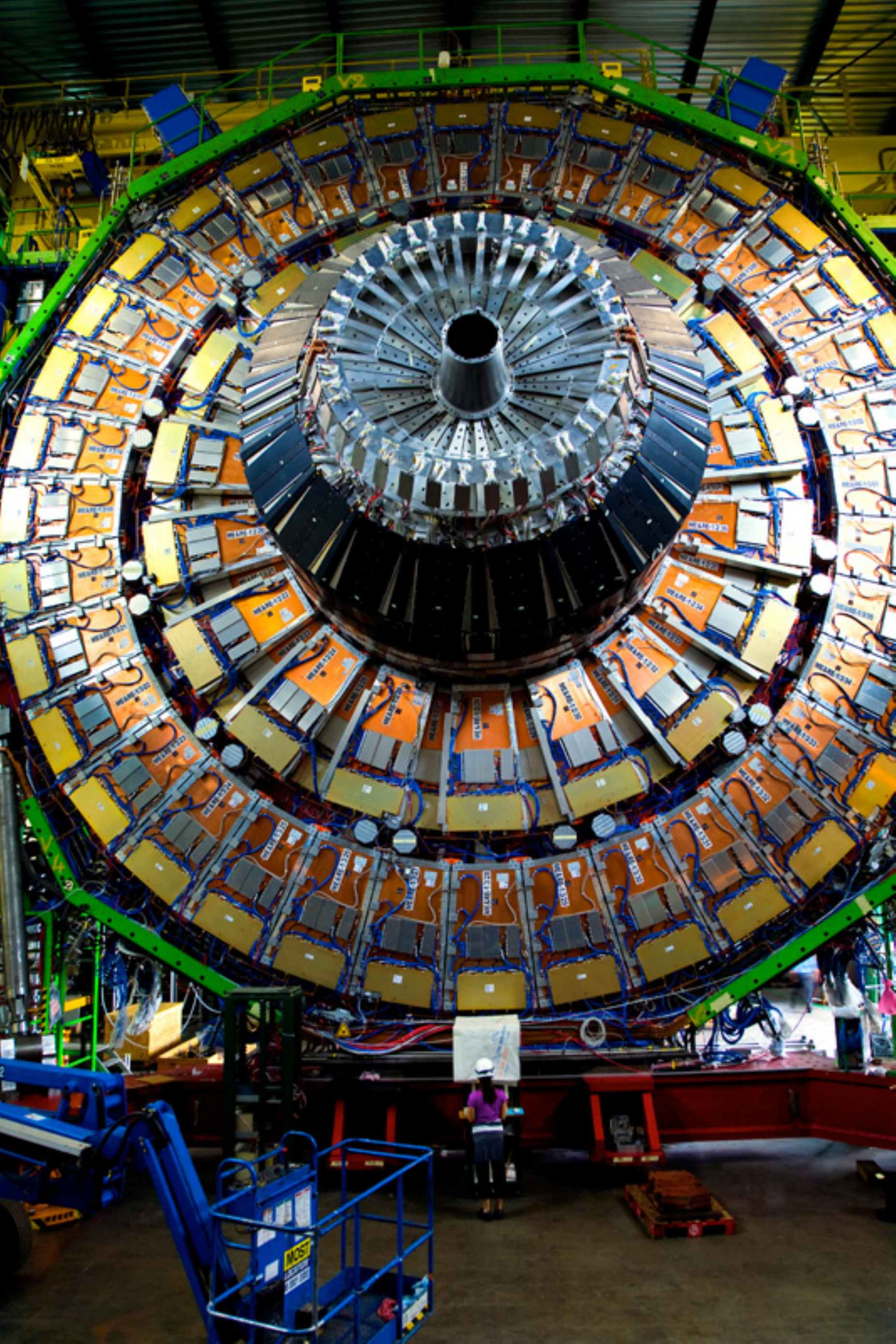}
 \caption{Photograph of the endcap wheel of the CMS detector, from \cite{NYTHiggs}.} \label{fig:CMSfront} 
\end{center} 
\end{figure} 

The New York Times articles on the Higgs boson included impressive pictures of the CMS detector, including the picture shown in Fig.~\ref{fig:CMSfront}~\cite{NYTHiggs}.  This endcap detector is one of the most visually imposing components of CMS and often appears in news reports.   It must be noted, then,
 that  every detector element seen in this picture was actually produced in the United States.  The orange detector elements are the Cathode Strip Detectors for muon tracking in the CMS endcaps.  These elements were constructed at the University of Florida and at Fermilab.  The technology used was invented in the U.S., originally to be applied to experiments at the Superconducting
 SuperCollider.  The large central black cylinder is the forward hadron calorimeter.   The detector elements are based on scintillator and wavelength-shifting fiber.   These elements were produced at Fermilab and reflect technology developed by that laboratory.    The design of these two systems was done by international consortia of laboratories; however, the U.S.  partners had the leading role.   Finally, the silver circular elements form the endcap of the   electromagnetic calorimeter.  The beautiful crystal calorimeter of CMS was mainly contributed by European and Chinese groups.  However, what is shown in the picture is the laser calibration system, which was designed and built at Caltech.  The final assembly of the module was done at CERN.  However, almost all of the  money and effort needed to produce this component of the CMS detector was spent in the U.S.

\subsubsection{ATLAS pixel tracker}

The construction of the ATLAS pixel tracker illustrates a more complex interplay among U.S.  national laboratories, U.S.  universities, and foreign groups.  The construction of this detector required, first, the fabrication of the  primary silicon sensors and their mounting structure, second, the design of a readout system, and third, overall mechanical integration and the supply of cooling and other services.  The responsibility for these various elements was divided among many institutions in the U.S., Germany, Italy, and France.  The ATLAS group at LBNL coordinated the U.S.  contributions.

The sensors are arranged in three cylindrical and four disk-shaped layers.  The European groups supplied the cylindrical layers, and the U.S.  took responsibility for the disks.  The disks were actually assembled at LBNL.

LBNL also took responsibility for the most complex component, the readout chip.   A breakthrough was the realization by the LBNL group that a technology recently introduced by IBM,  0.25 micron CMOS, was intrinsically radiation-hard because of very thin gate-oxides.  This capability was tested by an international collaboration of groups from both ATLAS and CMS.   This solution, involving a technology already industrialized in the U.S., allowed the full set of readout chips to be fabricated at a reasonable cost. This was essential to control the overall cost of the pixel detector.

  A host of other technologies needed to be applied to communicate with this chip.  These included a flex connect developed at the University of Oklahoma, an optical readout provided by Ohio State University, and readout drivers provided by the University of Wisconsin. The mechanical support engineering for the discs was a collaboration of LBNL with a small R\&D-oriented company in the U.S., Hytec.

In some of the case illustrated above, technologies already available in the U.S.  were the solutions of choice, in others, technologies developed by other international partners played a key role.  In many cases, ideas for novel detectors that originated in the U.S., including 3D detectors and diamond detectors, received the large majority of their funding for development from outside the U.S.

\subsection{Contributions to Accelerators}

Offshore accelerator projects also  benefit the domestic program in many diverse ways: scientifically,  economically,  technologically and socially.   We first discuss these benefits in general, and then give more detail for two  specific cases,  the LHC upgrade projects and the International
Linear Collider (ILC) R\&D program.

\subsubsection{Accelerator collaboration in overseas projects}

Cooperation in accelerator projects differs from cooperation in particle experiments in several ways.   These reflect the fact that accelerator projects are larger in scale and also that the accelerator science community is smaller.  The large scale of accelerators implies that these are the cost drivers for any advance in the power of high-energy physics facilities.   This gives increased importance to cost-sharing and the use of economies of scale.   The small size of the accelerator physics community makes it easier for international colleagues to work together to share key ideas and devote effort to common problems.  Both forces call for efforts in accelerator design and construction to become more internationalized.

\begin{table}[tp]
\centering
\begin{tabular}{lr}
\hline\hline
Source			& Equipment \& Personnel				\\ 
\hline
FNAL, BNL, and LBNL design and construction	& \$110M  	\\
Components purchased from U.S. industry  & \$90M		\\
\hline
Total									& \$200M	\\
\hline\hline
\end{tabular}
\caption{U.S. DOE resources spent on the  LHC accelerator complex~\cite{resources}.}
\label{tab:acc} 
\end{table}

There are many examples in which pooling of resources in accelerator development has led to results much improved over what individual regions could have achieved in isolation.  The LHC gives an example of this.    The large dipole magnets that fill the 27~km LHC ring are a European contribution.  However, the LHC also requires numerous specialized magnets for particular purposes.  Creating a high-luminosity interaction region requires high-field quadrupole magnets with specialized properties.  These were designed by U.S.  national labs in collaboration with KEK in Japan and provided in-kind to CERN.  The division of the total cost between international partners also allowed flexibility in assembling the total resources for the project in a timely way.

The U.S. contribution to this project engaged the this country's substantial  and talented beams physics community. Most of the U.S. accelerator 
physicists are  at national laboratories and, here also, the national laboratories led the effort.  This 
involved both ground-breaking accelerator R\&D and also large-scale fabrication of 
accelerator magnets. A wide variety of special-purpose magnets were designed, produced, installed, and maintained in the LHC tunnel by U.S. physicists and engineers. Table~\ref{tab:acc} shows the U.S. contribution to this project. All funds accounted here were spent on U.S. soil.

Participation in a collaboration of this type benefits the U.S.  domestic accelerator program in three ways---in making important technical resources available, in technology sharing and transfer, and in the recruitment of young scientists.  For the first of these goals, accelerator development requires test beams and specialized tooling for the manufacture of specific components.  An international partnership allows these facilities to be distributed over different regions, so that there is no need for duplication of effort, and motivates each nation or region to develop a complex of resources to fulfill its obligations to the collaboration.  These can be sources of  unique, world-leading expertise.   We will give examples below.

Development of close ties with international partners and common research programs facilitates  technology sharing and transfer.   Each accelerator project has specific, well-defined objectives that must find technical solutions.    Through collaboration, accelerator scientists in one region can draw on expertise and technology that would not otherwise be available to them to solve these problems.  This expands the entire technical base of the field.   An example of this is the development of superconducting technology for RF cavities and for magnets.   The design teams of the LHC and the ILC shared diverse specialized knowledge and infrastructure to improve both of these  technologies and to promote the production of components of both types at industrial scale.

Finally, the engagement of the U.S.  accelerator community in a diverse group of projects -- and, especially, those devoted to creating beams of the highest possible energies -- draws in students who will devote themselves to this technology and become its stewards in the future.   The accelerator community's social organization, its  tight-knit nature and international connections, provide a further incentive for students to join. The renewal of the accelerator community through the training of students is of course a necessity for the continuation of the U.S.  domestic accelerator program benefiting many fields of science.

\subsubsection{The LHC luminosity upgrade}

The two fundamental enabling technologies for the next generation of projects in high-energy physics are superconducting RF cavities and superconducting magnets.    These technologies play a key role also in current and proposed projects for Nuclear Physics and Basic Energy Sciences.  In this section and the next, we describe how the development of these technologies is now being driven through international collaboration.

The U.S.  has had a significant involvement with the LHC accelerator program beginning at a very early stage.  During the initial construction phase, in collaboration with KEK, Fermilab designed and constructed the final focusing quadrupole triplets and the  cryogenic feedboxes for the CMS and ATLAS high luminosity interaction regions.  In addition, Brookhaven built and contributed the superconducting D1 separator dipoles near the interaction region.

Based on the success of these ad hoc collaborations, the U.S.  LHC Accelerator Research Program (LARP) was formed in 2003 to coordinate U.S.  contributions to the LHC accelerator.  The organization made numerous contributions to the initial operation of the LHC, primarily through instrumentation, and has also supported U.S.  personnel working on the LHC.  However, the primary focus of the program has been R\&D related to the  planned luminosity upgrade of the LHC, currently scheduled to begin in 2022.

A key problem for this upgrade is to decrease the transverse beam size by about a factor of four.   The most straightforward way to achieve this is to replace the existing  NbTi superconducting quadupole magnets with  larger aperture, higher gradient versions  based on the next-generation superconductor Nb$_3$Sn. The development of  this technology has been the primary focus of the LARP program.  At  the small values of $\beta^*$ required, the fact that the beams cross  at a nonzero angle significantly  reduces the luminosity, a decrement called the  ``hourglass effect''. This effect must be compensated  to realize the potential luminosity  gains.  Thus, a second major thrust of the LARP program has been  R\&D into ``crab  cavities,'' lateral deflecting cavities that cause the  bunches to collide head-on in spite of the fact that the beams cross  at an angle.   The LARP program also includes a number of other projects on such topics as  beam stability and beam safety and abort mechanisms. The two major research topics, especially, are key issues in accelerator science that will influence the design of all future proton accelerators.

The LARP program will also construct equipment that will be part of the eventual High-Luminosity LHC. The DOE has requested that LARP submit a plan for a series of deliverables to the LHC luminosity upgrade, with a total project cost of approximately \$250M.

The LARP magnet program is acknowledged to be several years ahead of efforts in other regions in terms of the development of Nb$_3$Sn for practical use in accelerators.  The LARP initiative in crab cavities exploits the substantial investment that the U.S.  has made over many years in superconducting RF cavities. To date, all of the prototypes for crab cavities, including one designed in the UK, have been fabricated in the U.S. 

The overall picture is similar to the one we saw above  in our discussion of contributions to experiments. The support of the LARP program is money spent in the U.S.  for the development of key technologies in which the U.S.  is a leader. It will preserve our leading position in these areas both as a contribution to the overall High Luminosity LHC project and as a resource for the future of our domestic accelerator program.

\subsubsection{The ILC R\&D program}

In contrast to the LARP program, the ILC R\&D program provides an example in which the U.S.  was able to acquire a key technology through international collaboration.

The ILC R\&D program started in 2007 after the decision by ICFA that the next-generation linear electron-positron collider should be based on superconducting RF  technology. The program was established under the auspices of the Global Design Effort, a regional collaboration of Asia, Europe and the Americas. The principal technology goal was to demonstrate the viability of high accelerating gradients of   35~MV/m,  well beyond the state of the art at that time.

 Superconducting Radio Frequency (SRF) technology was original developed in the U.S.  at Cornell University and the Jefferson Laboratory.   Later, the R\&D center moved to DESY, where this technology was pursued for the TESLA program at DESY.  By the end of the 1990's, what remained in the U.S.  was little more than a lab-based curiosity, with minimal implementation in U.S.  projects.  The designs available for projects were  at quite low gradients.

The ILC GDE established an international cavity R\&D group, meeting monthly, to collect and share  the global expertise.  Initially, the program relied heavily on the knowledge from DESY.   This  allowed the U.S.  to renew and jump-start its domestic R\&D program. At the same time, a U.S.  industrial capability for cavity production had to be build afresh, starting from limited production capabilities in-house at Cornell and JLab.  Fermilab soon joined these two labs, and the U.S.  program began to grow quite rapidly.  New infrastructure was built at both Fermilab/ANL and at JLab, along lines guided by the DESY program.  The global experience base in this difficult technology was freely shared among the GDE participants, allowing both the U.S.  and Asia to  greatly reduce the necessary learning curve.  Within several years the basic technology was transferred to U.S.  industry.  At this time,  all superconducting  cavities for the ILC program are now produced in industry.

New processing techniques were developed both in Japan and the U.S.   When these were adopted globally,  they resulted in successful proof-of-principle cavities that  achieved the desired technical gradient goal of 35 MV/m.  Continued close collaboration with U.S.  industry to eliminate failure modes resulted in a final  yield in excess of 90\% for cavities meeting the high gradient goal.

The development of an SRF production program in the U.S.  required cavity processing and test facilities. Cavity processing facilities were created at Fermilab, ANL, JLab, and Cornell.  Vertical testing of cavities and horizontal testing of cryomodules is now possible at both Fermilab and JLab.   One U.S.  industrial cavity vendor is now certified through these tests as meeting the ILC technical  requirements.   A second vendor is now online though not yet certified.

The ILC R\&D program also includes examples in which research into  technologies in which the U.S.  was a leader was supported, including engineeering designs that made use of these technologies.    An example is the modulator system  that feeds RF power to the ILC cavities.   Significant advances in modulator design with solid-state components were made at SLAC, in collaboration with industry, and these are now incorporated into the overall ILC design.

\subsubsection{Accelerator innovation: conclusions }

In this section, we have discussed the many benefits that collaboration in world-class accelerator projects in other regions give to the U.S.  domestic program in accelerator science. These collaborations both support key technologies in which the U.S.  has a leading role and allow the U.S.  to take advantage of technology developments no matter where in the world they originate.  Programs at the national laboratories based on these technologies inevitably lead to U.S.  industrial spin-offs which enhance national capability and international competitiveness. HEP technology development helps not just the national program  but also other DOE programs.

The national SRF effort was completely rejuvenated by the ILC
program. Now superconducting RF is used in the major circular X-rays
sources NSLS and ALS and will be included in the design of new X-ray
sources based on free-electron lasers.   This program also provided
the capability for the JLab 12-GeV upgrade and forms the basis
 for the Fermilab Project X proposal for proton beams of unparalleled intensity.

While the ILC program successfully re-transplanted superconducting RF technology back into  the U.S., the LARP program built on existing U.S.  expertise in magnet technology to develop world leading next generation capabilities based on the superconductor Nb$_3$Sn.  The increase in magnetic fields and forces together with the brittle nature of the Nb$_3$Sn superconductor  make these magnets a major challenge.  The application of this technology to the LHC luminosity upgrade focused the U.S.   efforts and resulted in accelerator design options not available otherwise.  U.S.  industry is now a leader in fabricating this type of superconductor.

In recent years the U.S.  HEP budgets have fallen substantially in real terms.  The current CERN budget of 1.2B CHF is almost double the size of the total U.S.  program and represents only about 50\% of the EU particle physics program.  Asia is also beginning to ramp up its investment in high-energy accelerators, a trend that will become especially strong if Japan will host the ILC.  The examples we have given here illustrate the great opportunity for U.S.  accelerator physics that is available in collaborating with these programs and the benefits that will come from that collaboration.

\section{Particle Physics Imagination-Transfer}

The nearly 1500 U.S. scientists who make the LHC detectors work,
 who painstakingly characterize
the behavior and response of these detectors, and who analyze the 
data taken by these detectors for publishable physics  results, mostly 
work at U.S. universities. Most of these scientists are students or
postdocs for whom work on the LHC experiments is also part of their
scientific training.   Most of them will eventually  go on to jobs in
industries outside of basic research. While statistics are hard to
gather, a rule of thumb is that only  half of the graduate students go
on to postdocs in particle physics, and only half of those postdocs
end up permanently employed in particle physics in some capacity. 
These fractions have been lower in the past few  years.  So
the bulk of the people with our unusual technical training take their
experience and skills and apply
 them in some other areas of  the U.S. economy.

In this section, we will discuss U.S.  university and laboratory participation in the analysis of the LHC data.   As with the contributions to detectors, it is important to understand what work members of U.S.  collaborating groups actually carry out.  We will discuss this through some specific examples.

\subsection{Contributions to Experimental Analysis}

A physics result published by a large High Energy Physics
collaboration---for example, the mass of a particle or the value of a
cross section---is a composite of many lower-level analyses.  
The difficult aspects of 
an analysis typically involve validating that certain electronic
signals actually correspond to 
known particles in reliable ways and calibrating the corresponding
determination of particle 
properties.  The analysis convener and group members  coordinate
multiple studies and 
oversee their assembly into the final scientific result.

\subsubsection{An analysis contributor -- 1}

An example of a graduate student experience in a large collaboration
is given by Ryan 
Rios' Ph.D. research for the ATLAS collaboration.  Rios was a student
of Ryszard 
Stroynowski and a member of the ATLAS group at Southern Methodist
University, 
from 2007 to 2012. 

The first part of Rios' Ph.D. work dealt with the instrumentation of
the ATLAS detector in its commissioning phase.   SMU had
responsibility for the operation of the ATLAS Liquid Argon
Calorimeter.  Rios developed the monitoring panel for the trigger and
data acquisition 
software interface in the ATLAS control room. He wrote the crate
control and temperature monitoring sections and all of the interfaces
to the Relation Database and the Data Access Library.   As he moved
his 
orientation away from hardware to data analysis, he continued to
improve 
these systems, adapting the control panel to every subsequent TDAQ
software release, working as an  expert shifter for the Liquid Argon
system.  Eventually, he became a 
Shift Leader and also an on-call expert for the Liquid Argon 
on-line system.

Rios' detailed knowledge of the Liquid Argon system made it natural
for 
him to begin a data analysis topic in which electron identification
played a major role. In 2009,  Rios started to work on the search for
the Higgs boson decay to four electrons. 
Rios concentrated his effort on trying to maximize the electron
identification efficiency in the context of Higgs search.  This
project was done in collaboration with the SMU postdoc Julia Hoffman,
under the aegis of the ATLAS Higgs working group.   Most of the
background estimates and systematic uncertainties shown in December
2011 were based on his work.  Rios 
organized an ATLAS workshop in February 2012 that defined the final signal selection criteria and background estimates used for the four lepton mode in the  July 2012 Higgs discovery paper.

The work on this project was highly technically demanding, but it also required skills in communication and organization of technical material. In alternation with Hoffman, Rios gave numerous presentations to Higgs boson and electron/photon identification working groups at all levels in ATLAS.

Rios received his Ph.D. at SMU in May 2012 and stayed on with the group to work on  the July 2012 Higgs boson discovery paper.    However, after this, he decided to leave particle physics to pursue opportunities in industry.  He took a job in Houston with Lockheed Martin Space Systems, working on controls and data flow for communication with   the International Space Station.   In a similar way, the typical trajectory of graduate students in experimental high energy  physics is to bring the high level analytic and communication skills that their work has developed to applications in U.S.  high-technology industry.

\subsubsection{An analysis contributor -- 2}

Another example with a longer-term involvement in large
collaborations is that of Emmanuel Strauss.  
Strauss obtained his Ph.D. from SUNY Stony Brook in 2009,
analyzing data from the D\O\ experiment at the 
Tevatron collider at Fermilab. He then spent four
years as a postdoc with SLAC National Accelerator Laboratory working
on the
ATLAS experiment. 

At Stony Brook, Strauss worked with Profs. 
John Hobbs and Paul Grannis on the search for $ZZ$ production and 
contributed to the first observation of this rare process at a hadron collider.
As a part of this analysis, Strauss developed a new experimental
variable,
originally proposed by the University of Manchester's Terry Wyatt,
 to reduce the impact
of mismeasurement in the calculation of missing transverse energy.
Strauss also performed the statistical interpretation of the final 
results.
After the completion of this work, Strauss
contributed to the search for the Higgs boson decaying to two
b-quarks, produced in association with a Z 
boson decaying to a pair of leptons.  He developed  new analysis
techniques to improve the resolution of the measured quantities, 
and he also studied and applied machine 
learning methods to  enhance the sensitivity of the analysis.  His 
contributions were
 crucial for the evidence presented later by D\O\ for the Higgs boson.

After obtaining his Ph.D., Strauss joined the SLAC National Accelerator
Laboratory as a 
Research Associate in the ATLAS group. He was stationed at CERN for
three years.  During this time, he participated in new physics
searches with the LHC data.  He was also deeply involved with the
Trigger/Data Acquisition 
system. He became a Trigger Expert, and subsequently Trigger
Operations Manager.
 In this role, he was tasked with making any operational decisions 
that required immediate action concerning the modifications of 
algorithms and menus. Emanuel and his co-manager, Frank Winklmeier, 
supervised a rotating pool of on-call experts, and provided feedback 
to the trigger management in formulating future budget 
proposals for ATLAS hardware, the allocation of resources,
 and the ongoing optimization of service tasks in an organization
with many international partners.

After returning to SLAC,  Strauss lead a small of team of postdocs and 
graduate students pioneering the use of techniques from the field of 
computer vision for the classification of hadronic jet activity in the 
ATLAS calorimeter. This novel  approach
 treats calorimeter towers as the pixels of a 
picture, to which image analysis techniques can then be applied.

Shortly after the end of his fourth year as a postdoc, Strauss left 
particle physics to take a position as a Senior Data Scientist in 
AT\&T's Big Data division. Contributing creatively 
to data science in a corporate 
environment requires a wide variety of skills.  Few fields provide
as well as high-energy collider physics 
the breadth of relevant experience. This includes not only 
challenging
technical analysis but also the experience of writing public 
code in collaboration with many other people, and of managing
an operation with pressing demands that must be met in real 
time.  Strauss' experience in all of these areas
 made him a highly desirable candidate for his current position.

\subsubsection{An analysis leader}

To discuss the leadership roles in physics analysis, one's first reaction might be to discuss the highest level.  The U.S.  is well represented here; in CMS, both the current spokesman, Joseph Incandela, and the current Physics Coordinator, Greg Landsberg, are professors at U.S.  universities.  But these are CEO-type positions, requiring constant travel, for which the occupant's university base is almost irrelevant.  A more illustrative story is found at the next level down in the hierarchy, in the coordination of a specific physics group.

For the years 2009--2010, Prof. Jeffrey Richman of the University of California, Santa Barbara, led the CMS Supersymmetry group, in collaboration with Alexander Tapper of University College, London.    During this period, Richman made about one trip per month to Geneva and spent one full summer at CERN.  However, the bulk of his coordination work was done from his home in Santa Barbara.

The role of a physics topical coordinator is to shepherd a large number of projects like that of Ryan Rios described above, and to coordinate their assembly into concrete physics results for publication.   Richman and Tapper developed their overall strategies through a series of CMS-wide workshops. These established working groups on the key measurement topics. This, however, was only the beginning of a long-term engagement.  Carrying out their  plan required almost daily video meetings, which typically began between 5am and 8am Pacific time. Richman also allotted time for  informal discussions with the many graduate students and postdoctoral fellow, located all over the world, who were actually working directly with the data.

Fortunately, in the 21st century, we have communication tools  that allow us to directly interact with colleagues in distant parts of the world, and to host international meetings at all hours.  CERN has been instrumental in supporting many of these services for the benefit of its collaborations.   This was true in 1990, when the problem of communicating internationally within particle physics experiments led to the invention of the World-Wide Web.   CERN and the HEP community continue on the forefront today.

Richman's example---and the manifest success of CMS is carrying out very deep searches for supersymmetric particles using the early LHC data---demonstrate that there is no obstacle to providing intellectual leadership to a physics experiment located in Europe from a residence nine time zones away.   Some travel and personal flexibility is of course required.   This is a small price to pay to work with the enormous intellectual resources that a global experimental collaboration provides.

\subsection{Assisting Universities' International Outreach}

Many universities are now interested in providing opportunities for
their students to work in international settings.   A research
opportunity with a large international experiment provides an
excellent setting for undergraduates to learn a wide variety of
skills. 
These projects are technical, but they also require collaboration and 
communication, by internet or in person, with people from different nations.

Brown University is an example of a university for which strategies for global engagement and international visibility have recently become  one of the top priorities. What university administrators are looking for goes beyond the traditional semester-long exchange programs for undergraduates. Their concern is that students should understand the diversity of cultures with which they will come into contact in their careers and be able to work comfortably  in social and business settings in an increasingly interconnected world.  Thus, they emphasize collaborations, exchange programs, internships, and partnerships involving  faculty, staff, and students through a formal Internationalization Initiative.  At its inauguration,  former President Ruth J. Simmons stated, ``Internationalization requires a new way of thinking  across the campus about ourselves and our place in the world."  

Brown University has made a major effort to engage in  research in the international arena and Brown has explicitly recognized that undergraduate participation in the LHC collaborations serves these goals. The university  supports undergraduate students with summer stipends to conduct research abroad. In the summer of 2013, the Brown CMS group hosted three undergraduates at CERN and one at DESY, who are contributing to upgrade R\&D for the hadron calorimeter and for the silicon tracker. The student proposals for LHC projects were recognized not only for their academic goals, but also for the special opportunity and experience they would provide in learning how to navigate in a large collaborative global  setting.

At an event celebrating the Higgs boson discovery, Brown University President Christina Paxson stressed the value of supporting long-term research without immediate payoffs and praised the international collaborative effort of the Higgs boson search. President Paxson has singled out the involvement of students with the CMS group to alumni and  to many high level visitors  to the university. She has also called attention to this program at recruitment events,  as an example of the broad opportunities, including those with an international flavor,  easily accessible to students at Brown.

The Brown High Energy Physics group's involvement, expertise and insights with international collaborations  is valued by the university.  It is recognized by a number of appointments on the university's  International Affairs committee, which  develops and approves  funding decisions for proposed initiatives to bring international collaborators to work at Brown.   Under this initiative, Brown has brought undergraduate and graduate students from Singapore, Thailand, and Vietnam, with full external funding, to work with the CMS experimental team.

\section{The Role of Fermilab}

Many of the examples of collaboration discussed above involve the U.S.  national laboratories, which    
provide essential support to the LHC program and other overseas accelerator initiatives.  Among these, Fermilab has the largest U.S. program in particle physics and is the only single-purpose laboratory devoted
to this area of science.  This gives Fermilab a special role. 

Fermilab's current resources and expertise define two clear elements of its mission.   Fermilab's primary resource for high-energy physics is a proton beam of unmatched intensity.   Its primary mission, then, will be to use this beam for important high-energy physics measurements involving neutrinos and weak interactions. In precisely the manner we outlined above, Fermilab has thus become the global host for the most significant neutrino and charged lepton experiments in the world. Fermilab will build it, and others will come.

Fermilab has also become the steward for high intensity proton accelerator R\&D and by extension, a home for exploring focused practical applications for such industries as medicine, security, and nuclear waste transmutation. Its special role with the State of Illinois is ground-breaking and a tangible example of the technology and expertise that is always rooted in \emph{ab initio} scientific motivation.

However, Fermilab also plays an important role in the support of U.S.  participation in the LHC program.  Fermilab is the host of the U.S.  CMS collaboration and of the CMS U.S.  Tier 1 data center.  The LHC Physics Center (LPC) at Fermilab is the largest center of technical expertise for CMS outside Geneva.  Fermilab is also the host of LARP and provides a similar nucleus of expertise for the LHC accelerator.  We have given many examples above of components of the  LHC accelerator and the CMS detector that have been fabricated at Fermilab.

Fermilab has a rich history of contributions to particle physics experiments at the highest
energies, including the hosting of global experimental collaborations at the Tevatron.
In the current era in which the the highest energy colliders are hosted overseas,  Fermilab's resources and
expertise give it a key role in facilitating the contributions to U.S. universities to these global projects.
It is our hope that the staff and the director of Fermilab will see this as an important part of the 
laboratory's identity and an important contribution to its overall mission.

\section{Conclusion}

In this report, we have discussed the benefits to the U.S.  of  its participation in international collaborations hosted at accelerators abroad. We have emphasized that a correct understanding of how these collaborations are organized and supported teaches us that these benefits are similar to those from fundamental science experiments on a smaller scale, done entirely on U.S.  soil.  Our participation in these experiments supports technical innovation at the university or  laboratory level.   The money that is spent directly benefits U.S.  industry and the U.S.  economy.   Young scientists receive advanced training and, in most cases, bring their skills to applications in 
 the broader economy.

What relative emphasis should we give, then, to experiments here in the U.S.  or experiments abroad?  That, we feel, depends entirely on the scientific importance of those experiments.   This is what drives technical innovation.  We should not replace that criterion for another one, and certainly not for one that incorrectly represents how scientists actually work and contribute to large projects.

Today, many students and young physicists are excited by the discovery of the Higgs boson and the promise of other discoveries at the highest energy.  We must let them follow their dreams, and we must support them in doing so.  The benefits from the national support of basic research will return to us all just the same.

\Acknowledgements

We are grateful to many of our colleagues for discussion of these issues over the course
of the Snowmass study.  We are especially
grateful to Tiziano Camporesi, Abid Patwa, Jeffrey Richman, Emmanuel
Strauss, and Kelen Tuttle for providing
information and insights that had an important influence  on this presentation.
The preparation of this report was supported in part by  the U.S.  Department of Energy,
                     contract DE--AC02--76SF00515.


\begin{thebibliography}{99}



\bibitem{PPLB}
D. Overbye, {\it New York Times}, March 4, 2013.

\bibitem{LHCcost}
Lyndon R. Evans, ed, \emph{The Large Hadron Collider: A Marvel of Technology}, EFPL Press, Boca Raton, FL, 2009, p40.


\bibitem{countacc} \emph{Accelerators for America’s Future} U.S. Department of Energy, 2010, p5.

\bibitem{man70}  \emph{Accelerators and Beams, Tools of Discovery and Innovation} 4th edition, published by Division of Beam Physics of the American Physical Society, 2013, p27.

\bibitem{cernbudget} \emph{Final Budget of the Organization for the fifty-eighth financial year 2012} CERN, 2012, p13.

\bibitem{LHCauthors} Private communication with CMS and ATLAS managers.


\bibitem{resources} http://www.uslhc.us/



\bibitem{NYTHiggs}
D. Overbye, {\it New York Times}, March 5, 2013.

\end{thebibliography}
\end{document}